\documentclass[pdflatex,sn-nature]{sn-jnl}

\usepackage{graphicx}%
\usepackage{multirow}%
\usepackage{amsmath,amssymb,amsfonts}%
\usepackage{amsthm}%
\usepackage{mathrsfs}%
\usepackage[title]{appendix}%
\usepackage{xcolor}%
\usepackage{textcomp}%
\usepackage{manyfoot}%
\usepackage{booktabs}%
\usepackage{algorithm}%
\usepackage{algorithmicx}%
\usepackage{algpseudocode}%
\usepackage{listings}%
\usepackage{stfloats}
\usepackage[version=4]{mhchem}

\theoremstyle{thmstyleone}%
\theoremstyle{thmstyletwo}%

\theoremstyle{thmstylethree}%

\begin{document}

\title[Article Title]{Strong coupling of collective optical resonances in dielectric metasurfaces}


\author[1,2,3]{\fnm{Izzatjon} \sur{Allayarov}}

\author[4]{\fnm{Vittorio} \sur{Aita}}

\author[4,5]{\fnm{Diane J.} \sur{Roth}}

\author[4]{\fnm{Boaz} \sur{van Casteren}}

\author[4,6]{\fnm{Anton Yu.} \sur{Bykov}}

\author*[3,7]{\fnm{Andrey B.} \sur{Evlyukhin}}\email{evlyukhin@iqo.uni-hannover.de}

\author*[4]{\fnm{Anatoly V.} \sur{Zayats}}\email{anatoly.zayats@kcl.ac.uk}

\author*[1,2,3]{\fnm{Antonio} \sur{Cal{\`a} Lesina}}\email{antonio.calalesina@hot.uni-hannover.de}

\affil[1]{\orgdiv{Hannover Centre for Optical Technologies}, \orgname{Leibniz University Hannover}, \orgaddress{\street{Nienburger Str. 17}, \city{Hannover}, \postcode{30167}, \state{State}, \country{Germany}}}

\affil[2]{\orgdiv{Institute of Transport and Automation Technology}, \orgname{Leibniz University Hannover}, \orgaddress{\street{An der Universit{\"a}t 2}, \city{Garbsen}, \postcode{30823}, \country{Germany}}}

\affil[3]{\orgdiv{Cluster of Excellence PhoenixD}, \orgname{Leibniz University Hannover}, \orgaddress{\street{Welfengarten 1A}, \city{Hannover}, \postcode{30167}, \country{Germany}}}

\affil[4]{\orgdiv{Department of Physics and London Centre for Nanotechnology}, \orgname{King's College London}, \orgaddress{\street{Strand}, \city{London}, \postcode{WC2R 2LS}, \country{UK}}}

\affil[5]{\orgdiv{QinetiQ}, \orgname{Cody Technology Park}, \orgaddress{\street{Ively Road}, \city{Farnborough}, \postcode{GU14 0LS}, \country{UK}}}

\affil[6]{\orgdiv{NanoPhotonics Centre, Cavendish Laboratory, Department of Physics}, \orgname{University of Cambridge}, \orgaddress{\street{}, \city{Cambridge}, \postcode{CB3 0HE}, \country{UK}}}

\affil[7]{\orgdiv{Institute of Quantum Optics}, \orgname{Leibniz University Hannover}, \orgaddress{\street{Welfengarten 1}, \city{Hannover}, \postcode{30167}, \country{Germany}}}



\abstract{Dielectric metasurfaces can achieve strong light-matter interaction based on several types of collective (nonlocal) resonances, such as surface lattice resonances (SLRs) and quasi bound states in the continuum (quasi-BICs). Spectral selectivity, field enhancement, and high and controllable Q-factors make these resonances appealing for technological applications in lasing, sensing, nonlinear optics, and quantum photon sources.
An emerging challenge focuses on tailoring light-matter interaction via mode coupling and hybridization between the fundamental resonances of a metasurface. While strong coupling phenomena have been demonstrated between various resonant modes, the interplay between collective resonances of different natures has not been observed to date.  
Here, we theoretically, numerically, and experimentally demonstrate the onset of coupling and hybridization between symmetry-protected quasi-BICs and SLRs in a dielectric metasurface. We show the emergence of anticrossing (or Rabi splitting) in the strong coupling regime with suppression of reflection, observed under TE-polarised excitation, and the manifestation of an accidental BIC under TM-polarised illumination as a result of energy exchange between the participating collective resonances in the weak coupling regime. The first effect is accompanied by hybridized near fields of the modes. The observed coupling mechanisms can be controlled by modifying the angle of incidence, polarisation, and surrounding environment. This foundational study on the coupling and hybridization of collective resonances offers insights that can be leveraged for the design of metasurfaces with targeted quasi-aBIC and collective hybridized resonances. It could also open new possibilities to control the near fields associated with such resonances, with promising applications in tunable nanophotonics and light manipulation.}




\maketitle

\section{Introduction}

Optical metasurfaces supporting collective resonances with high quality factor (Q), such as surface lattice resonances (SLRs)~\cite{DuXiong2022} and quasi-bound states in the continuum (quasi-BICs)~\cite{Azzam2021,Kang2023,babicheva2024mie}, have been widely studied in recent years. 
With their ability to trap light for a relatively long time and enhance local field intensity, these resonances can boost nonlinear emission processes, including second and third harmonic generation~\cite{ Mobini2021,Koshelev2019}, control spontaneous parametric down-conversion to generate entangled photon pairs and complex quantum states~\cite{Checkova2022}, increase the sensitivity of optical sensors~\cite{Zhenchao2023}, and enable low-threshold lasing~\cite{Kang2023}.

While SLRs can be easily created in plasmonic metasurfaces based on dipolar interactions~\cite{Kravets2018}, as well as in dielectric metasurfaces~\cite{ Allayarov2023_APR}, achieving quasi-BICs in plasmonic systems requires tedious design engineering and minimization of intrinsic material losses~\cite{Sun2021,Azzam2018,Liang2020,Aigner2022,Liang2024}, making dielectric metasurfaces a better-suited platform for their realization~\cite{Kang2023,babicheva2024mie}. 
In general, the existence of BICs in dielectric metasurfaces (and their associated quasi-BIC resonances) can be understood via multipolar analysis~\cite{Sadrieva2019}. For example, symmetry-protected BICs (sBICs) are associated with multipoles of a meta-atom with zero radiation in the out-of-plane direction, for example electric dipoles. Their sensitivity to geometric and material parameters enables for their excitation and tunability via symmetry-breaking of the metasurface unit-cell~\cite{Evlyukhin2021, Mobini2021}, illumination at oblique incidence~\cite{murai2020bound,ustimenko2024resonances}, perturbation of the lattice~\cite{Yingying2024,Allayarov2024_PRB,Wang2023}, modification of the surrounding environment (which can include 2D materials)~\cite{Hu2024,Malek2024,Zhao2022}, or external stimuli such as electrical~\cite{Pura2023}, thermal~\cite{Zhaoyang2024}, and mechanical actuation~\cite{Huang2023}. 
Alternatively, radiating multipoles of the unit cell can lead to the realization of destructive interference conditions (for example, at a specific angle of incidence), generating quasi accidental BICs (quasi-aBICs).  
Due to their relatively broad spectrum, quasi-sBIC resonances have been demonstrated in several experiments~\cite{cui2018multiple,vabishchevich2018enhanced,liu2019high,xu2019dynamic,murai2020bound}. On the contrary, the extremely narrowband nature of quasi-aBIC resonances combined with the limited availability of analytical theories to predict their exact spectral and angular position~\cite{abujetas2022tailoring,Sadrieva2019}, has resulted in very few experimental demonstrations to date~\cite{sidorenko2021observation,Yingying2024}.

Although SLRs and quasi-sBICs can be explained using similar theoretical approaches (such as the dipole lattice sum~\cite{Allayarov2023_APR} and the ED-MD coupling~\cite{abujetas2020coupled}, respectively), the interplay between these two collective resonances has not yet been investigated and is the subject of this paper. We present a comprehensive theoretical, numerical, and experimental study of the coupling of such nonlocal modes in dielectric metasurfaces. We consider the dependence of collective resonances on the polarisation (TE and TM) and incidence angle of the incoming light, and the surrounding medium. Specifically, we demonstrate two effects deriving from the interaction between SLRs and quasi-sBICs: the emergence of strong coupling under TE polarisation, and the manifestation of an accidental BIC (aBIC) under TM polarisation as a result of near-field energy exchange between the two collective modes.

The investigation of strong coupling in metasurfaces is gaining increasing interest due to its potential to create hybrid photonic states, which can offer new possibilities for tailoring and enhancing light-matter interactions. Strong coupling involving quasi-BIC resonances has been observed with guided modes~\cite{Chen2024}, epsilon-near-zero modes~\cite{Yue2024}, excitons~\cite{Weber2023,Sortino2025}, as well as between adjacent quasi-BICs~\cite{Zhang2025,Biswas2024,Xie2024}. Strong coupling has also been reported between lattice resonances of a hybrid metal-dielectric metasurface~\cite{Oleynik2024}. However, strong coupling between collective resonances of different natures is to date unknown. Here, we theoretically and experimentally demonstrate strong coupling between quasi-sBICs and SLRs. In particular, under TE illumination, we observe the formation of hybrid nonlocal modes with suppression of reflection in the anticrossing spectral range due to strong coupling. We also demonstrate the possibility of tuning the coupling strength and bandwidth via the surrounding environment, angle of incidence, and polarisation.

Previous experimental demonstrations of aBICs describe their formation as the simple destructive interference between in-plane and out-of-plane multipolar components of the single meta-atom at specific angles of incidence with limited theoretical explanations of the underlying mechanisms~\cite{Yingying2024}. In this paper, we not only experimentally measure the aBIC and the associated quasi-aBIC hybrid modes, but also theoretically demonstrate the important role of the hybridization between SLR and quasi-sBIC collective resonances in the formation of the aBIC in a weak coupling regime.

We believe our comprehensive numerical, theoretical and experimental study provides new insights on nonlocal effects in dielectric metasurfaces, and can support the design (and potentially the reverse engineering) of metasurfaces with spectral and angular control of hybrid collective modes, therefore, expanding the possibilities for controlling and enhancing light-matter interactions.

\section{Results and discussion}

\subsection{Experimental and numerical results}

\begin{figure*}[!htb]
\centering
\includegraphics[width=1\linewidth]{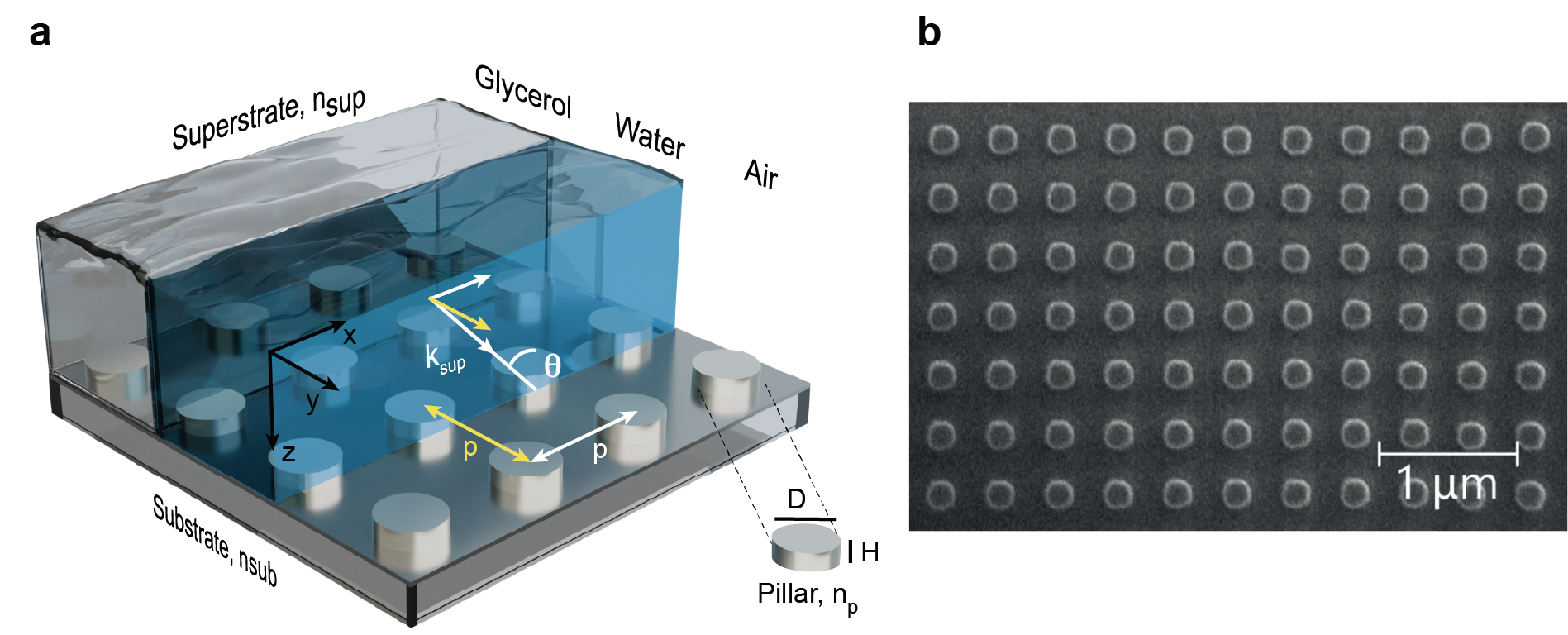}
\caption{(a) Schematics and (b) scanning electron microscope (SEM) image of the studied metasurface, consisting of a square array of polycrystalline Si disks with diameter D=220~nm, height H=100~nm, and period P=440~nm on a glass substrate with refractive index of $n_{\rm sub}$=1.45. Three different superstrates considered in this paper are also schematically shown in (a). The metasurface is illuminated from the top at an angle $\theta$ in the $xz$-plane, chosen as the plane of incidence. The white arrow labelled with $k_{\text{sup}}$ indicates the incident plane wave in a superstrate. Its polarisation is considered to be either transverse electric (TE, electric field along the $y$-axis) or transverse magnetic (TM, magnetic field along the $y$-axis).}
\label{fig:str}
\end{figure*}

The metasurface studied in this work consists of polycrystalline Si nanodisks with a diameter D = 220~nm, height H=100~nm, arranged in a square lattice with a period P=440~nm, with the global size of the array being 100$\times$100~$\mu$m$^2$~(Fig.~\ref{fig:str}). The nanodisks are obtained by electron beam lithography from a 100\,nm-thick Si film on a glass substrate ($n_{\rm sub}$ = 1.45). Under illumination at normal incidence, the optical response of such metasurface is determined by the SLR excitation and strongly dependent on the refractive index mismatch between the substrate and superstrate~\cite{Allayarov2023_APR}.

\begin{figure*}[t]
\centering
\includegraphics[width=1\linewidth]{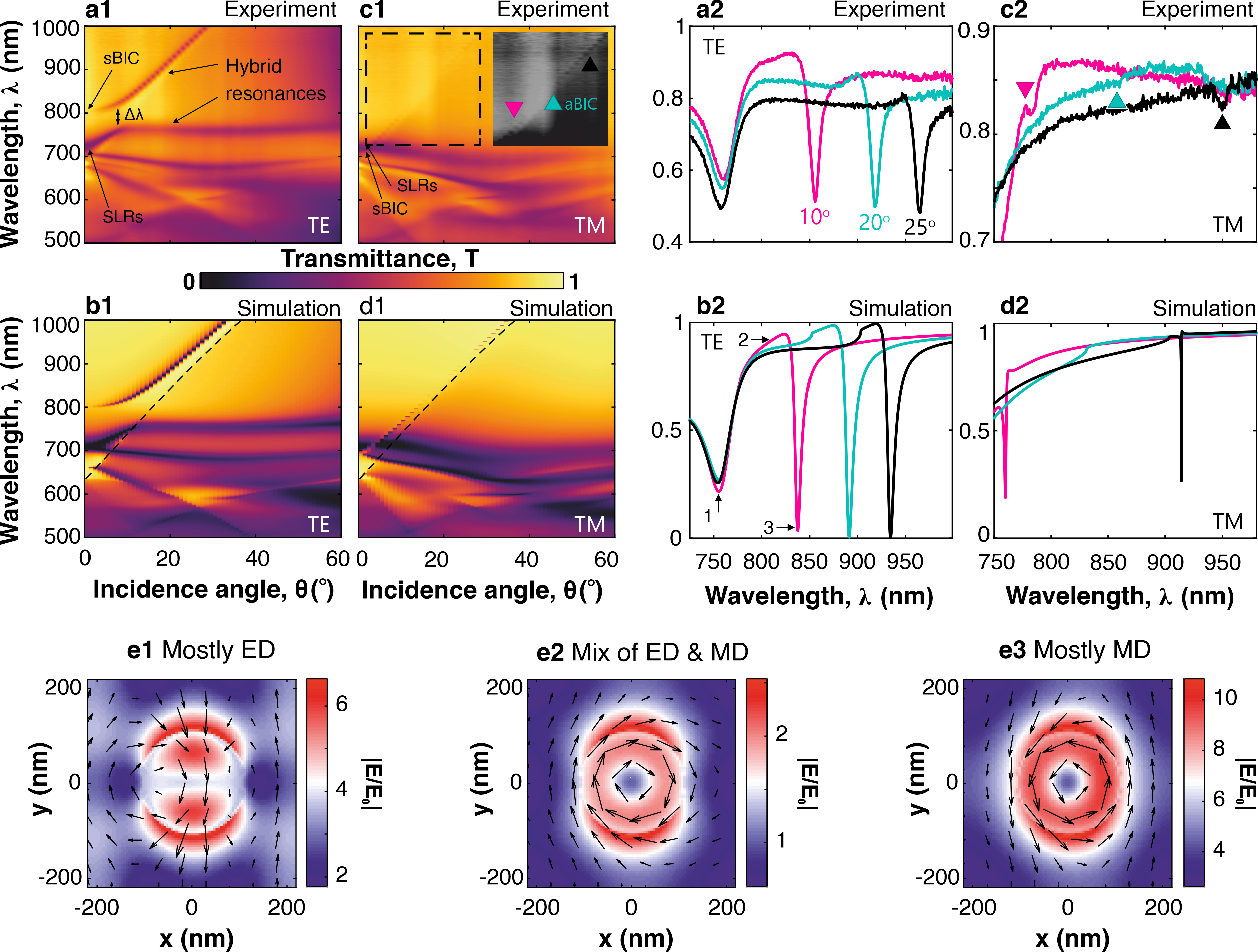}
\caption{(top row) Measured and (bottom row) simulated transmittance of the metasurface on a glass substrate ($n_{\rm sub}$=1.45) with index-matching glycerol ($n_{\rm sup}$=1.45) as a superstrate under (a,b) TE- and (c,d) TM-polarised illumination. Inset in panel (c1) shows the high-contrast grayscale image of the region enclosed in the dashed box. Dashed lines in (b1,d1) indicate the main order Rayleigh anomaly~\cite{ndao2018plasmonless} of the homogeneous environment. The resonances labelled by the black arrows in (a1,c1) are symmetry-protected BIC (sBIC) and accidental BIC (aBIC). (a1--d1) Transmittance dispersion with respect to incidence polar angle $\theta$. (a2--d2) Transmittance spectra at the selected incidence angles: (pink) $\theta = 10^\circ$, (teal) $\theta = 20^\circ$, and (black) $\theta = 25^\circ$. The values of the wavelength $\lambda$ are indicated for vacuum. In (a1) $\Delta \lambda$ indicates the anticrossing region (or Rabi splitting), whose width is determined by the coupling strength. (e1-e3) Electric field magnitude (colour scale) and direction (black arrows) distributions for the metasurface unit cell at the spectral positions indicated in (b2) for $\theta=10^\circ$.
}
\label{fig:t_g_exp}
\end{figure*}

In a uniform environment, achieved by submerging the metasurface in glycerol ($n_{\rm sup} = n_{\rm sub}$ = 1.45), the transmission spectra measured under TE-polarised illumination show a complex system of resonances, with the transmittance minima exhibiting a strong wavelength dependence on the angle of incidence, in excellent agreement with numerical modeling (Fig.~\ref{fig:t_g_exp}). 
Most of the resonances in Fig.~\ref{fig:t_g_exp}(a1) evolve from the features observed at normal incidence, such as sBIC and SLR (the origin of the latter is discussed in detail in Ref.~\cite{Allayarov2023_APR}).
The SLR resonance observed at normal incidence at a wavelength of $\lambda\approx$ 700 nm splits into two spectrally narrow (FWHM~$\approx$ 10~nm) and well-separated (60--80~nm) features that show a relatively small chromatic dispersion with the angle of incidence. 
An additional feature of low transmission arises at $\lambda\approx$ 800~nm, as soon as the angle of incidence becomes nonzero, which red-shifts as $\theta$ increases, asymptotically approaching the main order Rayleigh anomaly (RA, representing the condition for which a diffraction order of a periodic array becomes an evanescent wave propagating in the plane of the array) of the homogeneous environment (Fig.~\ref{fig:t_g_exp}(b1)). 
This feature is a quasi-sBIC, which shows an experimental quality factor of approximately 85 (theoretical Q~$\approx 100$). The emergence of a spectral band $\Delta \lambda$ with suppressed reflection (Fig.~\ref{fig:t_g_exp}(a1,a2)) due to the anticrossing of the SLR and quasi-sBIC features suggests the presence of strong coupling between these two collective resonances, as demonstrated in the next section. The value of the Rabi splitting energy experimentally measured is $\Delta \hbar \omega \approx$ 130~meV (corresponding to $\Delta \lambda\approx$ 65~nm). From the evolution of the electric field distributions in the anticrossing region, the resonances with the main contribution of (1) SLR, (3) quasi-sBIC, and (2) a hybridized mode between them can be identified (Fig.~\ref{fig:t_g_exp}(e1-e3)). From the comparison of the electric field values, it is also evident that, under resonance conditions (Fig.~\ref{fig:t_g_exp}(e1,e3)), the near field is amplified several times compared to the spectral region between resonances (Fig.~\ref{fig:t_g_exp}(e2)). This behavior is characteristic of the strong coupling regime and is due to the redistribution of energy between the two hybrid modes and their weak spectral overlap.

We also note the presence of additional high-order Rayleigh anomalies for wavelengths shorter than $\approx$650 nm at $\theta=0^\circ$ (Fig.~\ref{fig:t_g_exp} and Fig.~S2). However, they have a minimal influence on the effects considered here, and a detailed theoretical analysis of these diffractive contributions is beyond the scope of this paper.

In the case of TM-polarised illumination, the transmission shows a similar response as a function of wavelength and angle of incidence, and also excellent agreement between experiments and simulations (Fig.~\ref{fig:t_g_exp}(c1,d1)). However, in this case, no anticrossing is observed as the spectral gap between the SLR and sBIC modes is much smaller than the width of the resonances, and the modes are in a weak coupling regime. 
A narrow feature, observed along the main order RA for this polarisation, evolves differently with the incidence angle. It disappears at around $\theta =$ 20$^{\circ}$ and $\lambda\approx$ 850~nm and later on reappears for $\theta>$25$^\circ$ (inset in Fig.~\ref{fig:t_g_exp}(c1) and Fig.~\ref{fig:t_g_exp}(c2,d2)). 
As will be demonstrated later, this feature, known as aBIC, is the result of a weak coupling between SLR and sBIC modes.

\subsection{Coupled-dipole model and demonstration of strong coupling}
An isolated disk of the studied array (Fig.~\ref{fig:str}a) mainly exhibits ED and MD responses in the visible spectral range (Fig.~S1 of the Supporting Information) so that the spectral features observed experimentally can be interpreted by applying the coupled-dipole model (CMD)~\cite{allayarov2025analytical}. Although the model derived in Ref.~\cite{allayarov2025analytical} covers the most general case of meta-atoms with anisotropic polarizability, here we apply it to the sub-class of nanoparticles with in-plane isotropy. We note that approaches from Refs.~\cite{abujetas2018generalized,abujetas2020coupled,abujetas2022tailoring} using a different representation, can also be adopted. 

Within the CDM, the transmission $t$ (transmittance $T=|t|^2$) and the reflection $r$ (reflectance $R=|r|^2$) coefficients corresponding to the specular reflectance (the zero diffraction order)  of an infinite metasurface composed of identical particles and embedded in a homogeneous environment (with relative permittivity $\varepsilon_{\rm sur}$) can be written for the two TE and TM irradiation configurations as~\cite{allayarov2025analytical}

\begin{align}
 &r^{\rm TE} = \frac{{\rm i}k_{\rm sur}}{2S_{\rm L}\varepsilon_0\varepsilon_{\rm sur} E_0\cos\theta }\Big(\Big[p_y\!+\!\frac{\sin\theta}{v}m_z\Big]+\frac{\cos\theta}{v}m_x\Big), \label{rTE} \\
&t^{\rm TE} = 1+\frac{{\rm i}k_{\rm sur}}{2S_{\rm L}\varepsilon_0\varepsilon_{\rm sur} E_0\cos\theta }\Big(\Big[p_y\!+\!\frac{\sin\theta}{v}m_z\Big]-\frac{\cos\theta}{v}m_x\Big), \label{tTE} \\
&r^{\rm TM} = \frac{{\rm i}k_{\rm sur}}{2S_{\rm L}H_0\cos\theta }\Big(\Big[m_y-v\sin\theta p_z\Big]-v\cos\theta p_x\Big),\label{rTM} \\
&t^{\rm TM} = 1+\frac{{\rm i}k_{\rm sur}}{2S_{\rm L}H_0\cos\theta }\Big(\Big[m_y-v\sin\theta p_z\Big]+v\cos\theta p_x\Big),\label{tTM}
\end{align}
where $\varepsilon_0$ is the vacuum permittivity,  $S_{\rm L}$ is the lattice unit cell area, $k_{\rm sur}$ and $v$ are the wave number and speed of light in a surrounding homogeneous environment, respectively, $E_0$ and $H_0$ ($\varepsilon_0\varepsilon_{\rm sur}E_0=H_0/v$) are the electric and magnetic fields of the incident plane wave, respectively, at the position of the electric ${\bf p}=(p_x,p_y,p_z)$ and magnetic ${\bf m}=(m_x,m_y,m_z)$ dipole moments of the particle in the metasurface located at the origin of the selected Cartesian coordinate system. The incident electric (magnetic) field is along the $y$-axis in the case of TE (TM) polarisation. The nonzero dipole moments for TE polarisation are
\begin{align}
  &m_x=-\frac{H_0\cos\theta}{1/\alpha^{\rm m}_{\parallel}-S_x},\label{mx}\\
  &p_y=\frac{\left[1/\alpha^{\rm m}_{\perp}-S_z\right]-{\rm i}g_x\sin\theta}{[{1}/{\alpha^{\rm p}_{\parallel}}-S_y] \left[{1}/{\alpha^{\rm m}_{\perp}}-S_z\right]+g_x^2}  \varepsilon_0\varepsilon_{\rm s}E_0,\label{py}\\
  &m_z=\frac{[1/\alpha^{\rm p}_{\parallel}-S_y]\sin\theta-{\rm i}g_x}{[{1}/{\alpha^{\rm p}_{\parallel}}-S_y] \left[{1}/{\alpha^{\rm m}_{\perp}}-S_z\right]+g_x^2} H_0.\label{mz}
\end{align}
In the case of TM-polarised excitation, the nonzero dipole moments are
\begin{align}
  &p_x=\frac{\varepsilon_0\varepsilon_{\rm s}E_0\cos\theta}{1/\alpha^{\rm p}_{\parallel}-S_x},\label{px}\\
  &m_y=\frac{[1/\alpha^{\rm p}_{\perp}-S_z]-{\rm i}g_x\sin\theta}{\left[1/\alpha^{\rm m}_{\parallel}-S_y\right] \left[1/\alpha^{\rm p}_{\perp}-S_z\right]+g_x^2} H_0,\label{my}\\
  &p_z=\frac{-[1/\alpha^{\rm m}_{\parallel}-S_y]\sin\theta+{\rm i}g_x}{\left[1/\alpha^{\rm m}_{\parallel}-S_y\right] \left[1/\alpha^{\rm p}_{\perp}-S_z\right]+g_x^2} \varepsilon_0\varepsilon_{\rm s}E_0.\label{pz}
\end{align}
In the above equations, the in- and out-of-plane ED polarizability components of the isolated single particle are indicated as $\alpha_{\parallel}^{\rm p}$ and $\alpha_{\perp}^{\rm p}$, respectively, and the corresponding MD polarizabilties are indicated as  $\alpha_{\parallel}^{\rm m}$ and $\alpha_{\perp}^{\rm m}$, while $S_x=S_x(\theta)$, $S_y=S_y(\theta)$, $S_z=S_z(\theta)$ are the lattice sums corresponding to ED-ED and MD-MD interactions in the metasurface, and $g_x=g_x(\theta)$ is the ED-MD coupling parameter which vanishes at normal ($\theta=0^\circ$) incidence of the external wave. The explicit expressions of $S_x,S_y,S_z$ and $g_x$, and their dependence on wavelength and angle of incidence are presented in the Supporting Information (Eqs.~(S1--S4) and Fig.~S3). 

As follows from Eqs.~(\ref{mx}-\ref{mz}) for TE polarisation and Eqs.~(\ref{px}-\ref{pz}) for the TM polarisation, under normal illumination (when $g_x=0$) the ED-MD interaction disappears and the dipole moments are determined independently of each other:
 \begin{align}
     &\quad \qquad{\rm TE}\qquad\qquad\qquad\qquad {\rm TM}\nonumber\\
     &m_x=-\frac{H_0}{1/\alpha^{\rm m}_{\parallel}-S_x};\qquad p_x=\frac{\varepsilon_0\varepsilon_{\rm s}E_0}{1/\alpha^{\rm p}_{\parallel}-S_x};\\
&p_y=\frac{\varepsilon_0\varepsilon_{\rm s}E_0}{{1}/{\alpha^{\rm p}_{\parallel}}-S_y}; \qquad \quad 
m_y=\frac{H_0}{1/\alpha^{\rm m}_{\parallel}-S_y } ;\\
&m_z=0; \qquad \qquad \qquad \quad p_z=0.\label{zz}
 \end{align}
 Note that in this case the ED (MD) moments depend only on the ED (MD) polarizability of the isolated single particles.
 
Under oblique illumination ($\theta\ne 0^\circ$), an ED-MD interaction arises in the system: in the TE case of Eqs.~(\ref{py},\ref{mz}), $p_y$ and $m_z$, and in the TM case of Eqs.~(\ref{my},\ref{pz}), $m_y$ and $p_z$ are determined by both the ED and MD polarizabilities of isolated single particles. This in turn indicates a coupling between the corresponding ED and MD moments.
In Eqs.~(\ref{rTE}-\ref{tTM}), we have combined in square brackets, the dipole moments interacting with each other in the metasurface under the condition of oblique incidence of illuminating plane wave.  The main advantage of the analytical CDM formulation is that it enables a straightforward analysis of the coupling effect  between ED and MD. For example, in the case of TE-polarised illumination $p_y$ and $m_z$ (for the TM case, $m_y$ and $p_z$) are coupled via the parameter $g_x$ while the $m_x$ (for the TM case, $p_x$) component is independent. Within the CDM, the coupling parameter $g_x$ can be turned on and off in the simulations, allowing detailed studies on the specific role of the coupling. 
 
To demonstrate the applicability of the CDM, we compared the reflectance and transmittance spectra of the metasurface, with parameters as  in Fig.~\ref{fig:t_g_exp}, calculated using CDM and full-wave numerical simulations. These show excellent agreement with the numerical results (Fig.~S4), confirming its applicability. To clearly demonstrate the dipole coupling, in the following we focus on the reflective properties of the metasurface, which are determined only by the contribution of scattered waves. The same considerations govern the transmission.

\begin{figure*}[t]
\centering
\includegraphics[width=1\linewidth]{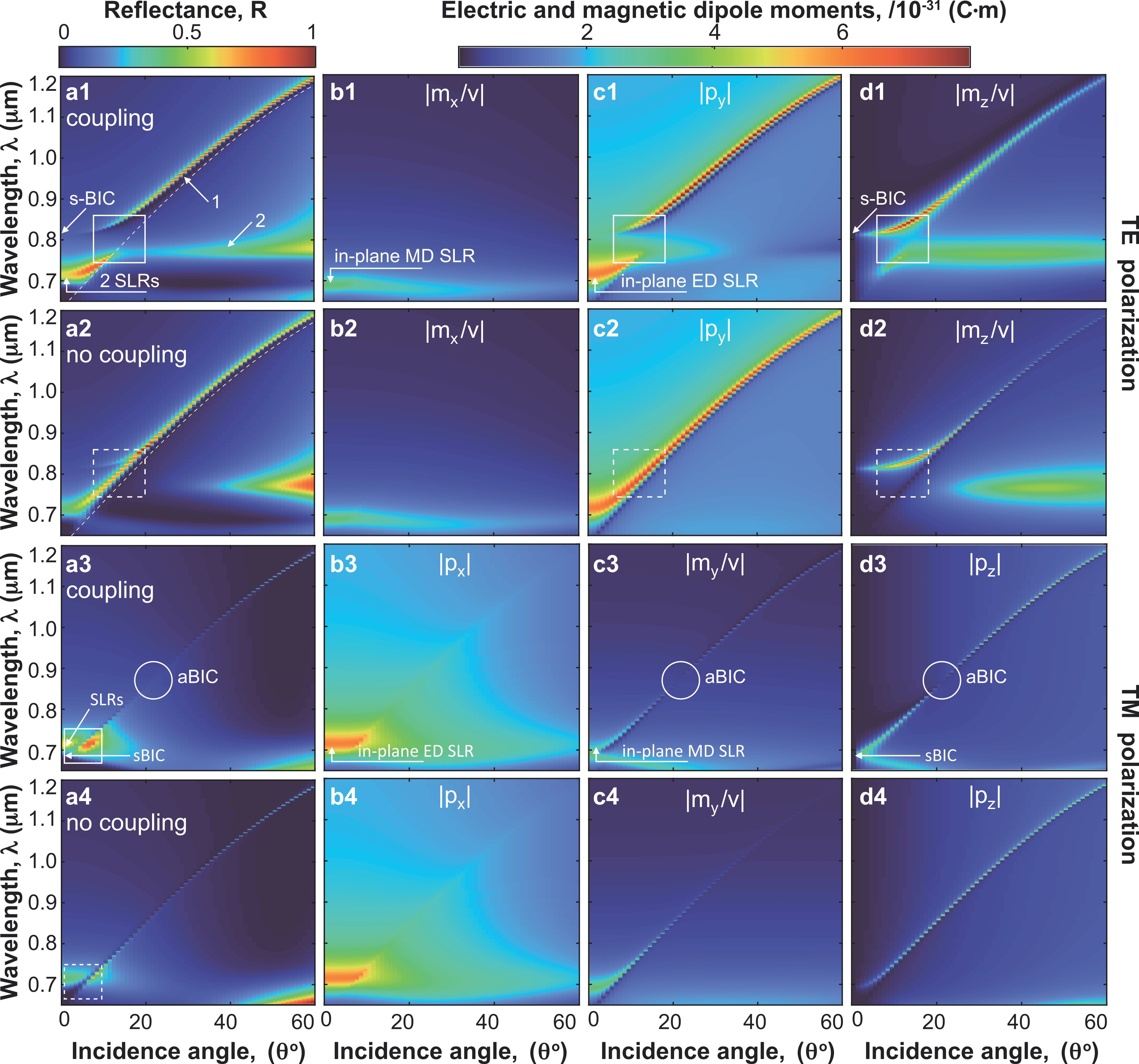}
\caption{Simulated (a) reflectance spectra and (b--d) contributing nonzero dipole moments of the metasurface (as indicated in the panel legends), calculated from the CDM for (1,2) TE and (3,4) TM polarisations for different incidence polar angles $\theta$ (1,3) with and (2,4) without the inclusion of the coupling term between the in- and out-of-plane dipolar components. Features labels and the parameters of the metasurface are as in Fig.~\ref{fig:t_g_exp}. The values of the wavelength $\lambda$ are indicated for vacuum.}
\label{fig:r_glycerin_num}
\end{figure*} 

As follows from the above equations describing CDM, at normal incidence ($\theta=0^\circ$), the out-of-plane components (${z}$-components) of the dipole moments do not contribute to the transmission or reflection and cannot couple to any other component of the dipole moments due to $g_x=0$. 
It is therefore impossible to excite the out-of-plane components of the dipole moments at normal incidence (Eqs.~(\ref{zz})) and they can only contribute to sBIC formation~\cite{Evlyukhin2021}. As a consequence, the resonant features, which are observed at normal incidence both in transmission and reflection are only related to the in-plane components ($x$, $y$) of the dipole moments, determining the excitation of the SLRs~\cite{Allayarov2023_APR}. 
Under oblique illumination, in-plane and out-of-plane components of the dipole moments can be excited, with the coupling being allowed  between ED and MD excitations. 
Due to the coupling, the resonant excitation of the out-of-plane dipole moment associated with the sBIC automatically influences the excitation of the in-plane dipole moment associated with the SLR, and vice versa.
At small incidence angles, the mutual influence between quasi-sBIC and SLR is small.
With the increase of the angle of incidence, the interaction between the dipole moments corresponding to the sBIC and the SLR increases and the system enters the regime of mode hybridization, leading to several exotic phenomena depending on polarisation, as explained below. Moreover, the formation of hybrid modes is accompanied by tailored near fields (Fig.~\ref{fig:t_g_exp}(e1-e3) and Fig.~S4), which can be useful in applications. 

Using the results of the CDM theory presented in Fig~\ref{fig:r_glycerin_num}, the experimental and simulated transmittance and reflectance spectra can be understood. 
First, we consider the spectra for TE-polarised illumination (Fig.~\ref{fig:r_glycerin_num}(a1--d1)).
At normal incidence, the metasurface reflectance spectra exhibit two overlapping SLRs around $\lambda=700$~nm: a weak in-plane MD SLR with $m_x$ and a strong in-plane ED one with $p_y$~\cite{Allayarov2023_APR}. Since
the spectral position of the SLR with $m_x$ is practically independent of the angle of incidence and does not contribute to ED-MD coupling (cf. Fig.~\ref{fig:r_glycerin_num}(b1) and Fig.~\ref{fig:r_glycerin_num}(b2)), we focus on the behavior of $m_z$ and $p_y$ dipoles.

A narrow resonance at around $\lambda=800$~nm for incidence angles $0^{\circ}<\theta<5^{\circ}$ (Fig.~\ref{fig:r_glycerin_num}(a1)) corresponds to the transformation of the nonradiating sBIC into the radiating quasi-sBIC, and is caused by the out-of-plane MD moment component $m_z$ (Fig.~S5(a1-d1))~\cite{murai2020bound,abujetas2020coupled,abujetas2022tailoring}. Notably, for small angles of incidence ($0^{\circ}<\theta<5^{\circ}$), the spectral gap between the quasi-sBIC and SLRs is larger than the width of these resonances themselves (Fig.~\ref{fig:r_glycerin_num}(a1)). Additionally, in this range of angles of incidence, the values of the dipole moments $m_z$ and $p_y$ determining the quasi-sBIC and SLR, respectively, are not influenced by coupling (cf. Fig.~\ref{fig:r_glycerin_num}(c1,d1) and Fig.~\ref{fig:r_glycerin_num}(c2,d2)). 

The situation changes considerably with increasing angle of incidence $\theta$. At angles $\theta>10^{\circ}$, the system enters a strong coupling regime between the quasi-sBIC with $m_z$ and the SLR with $p_y$, leading to the anticrossing effect (Fig.~\ref{fig:r_glycerin_num}(a1,c1,d1) and Fig.~S6(a1,b1)) 
with an anticrossing gap larger than the linewidth of the SLR and quasi-sBIC modes. According to the general strong coupling theory \cite{novotny2010strong}, an anticrossing effect results in two hybrid modes, each of which includes dipole moments from the two contributing modes.
In this case, two non-intersecting hybrid BIC-SLR modes are formed with participation of magnetic $m_z$ and electric $p_y$ dipoles in each hybrid mode. These modes do not intersect with each other due to the symmetry of the related fields. With the increase of the angle of incidence, and consequently with the increase of the coupling strength, the spectral gap between these modes increases in accordance with the general theory of strong coupling~\cite{novotny2010strong}. Note that if the coupling parameter is excluded from the CDM, no anticrossing effect occurs in the reflection spectrum (Fig.~\ref{fig:r_glycerin_num}(a2)).

Formally, using the contribution of the coupled dipoles to the reflection coefficient (Eq.~\ref{rTE}), an effective polarizability can be defined for the hybrid modes as
\begin{equation}\label{pH_TE}
    \alpha^{\rm HYB/TE}(\theta)=\frac{p_y+(m_z\sin\theta)/v}{\varepsilon_0\varepsilon_{\rm sur}E_0}\:.
\end{equation}
In this context, since the hybrid modes are formed by two resonant dipole contributions, which generally possess different quality factors, the resonance of the coupled system will exhibit a Fano profile (see, for example, the resonant band between the hybrid resonances in Fig~\ref{fig:r_glycerin_num}(a1), and the corresponding experimental and simulated spectra in Fig.~\ref{fig:t_g_exp}(a2,b2)). Note that the hybrid mode at longer wavelength in Fig~\ref{fig:r_glycerin_num}(a1) follows the Rayleigh anomaly on the diffraction-free side of the dispersion. This is due to the fact that, neglecting coupling between the modes, the quasi-sBIC (Fig.~\ref{fig:r_glycerin_num}(d2)) and SLR (Fig.~\ref{fig:r_glycerin_num}(c2)) are also formed in this region with increasing angle of incidence. To further prove the emergence of strong coupling, we evaluate the required conditions for the coupling coefficients~\cite{zhang2018photonic,cao2020normal,Weber2023}: 
\begin{align}
& c_1 = \Delta \omega / (\Delta \omega_{\rm SLR} + \Delta \omega_{\rm q-sBIC}) > 1, \\
& c_2 = [(\Delta \omega^2 + (\Delta \omega_{\rm SLR} - \Delta \omega_{\rm q-sBIC})^2)/(2(\Delta \omega_{\rm SLR}^2 + \Delta \omega_{\rm q-sBIC}^2))]^{1/2} > 1,
\end{align}
where $\Delta \omega = \omega_{\rm SLR} - \omega_{\rm q-sBIC}$ is the spectral gap between SLR and quasi-sBIC modes, $\Delta\omega_{\rm SLR}$ and $\Delta\omega_{\rm q-sBIC}$ are their FWHMs. For TE polarisation, these coefficients determined from simulation are $c_1=2.46$ and $c_2=2.26$, and $c_1=2.27$ and $c_2=2.17$ from the experiment. This confirms that strong coupling is achieved between SLR and quasi-sBIC modes.

For TM polarisation and in the range of small angles of incidence (Figs.~\ref{fig:r_glycerin_num}(a3--d3) and (a4--d4)), the results obtained are similar to the TE case, with the effective hybrid polarizability determined by 
\begin{equation}\label{pH_TM}
    \alpha^{\rm HYB/TM}(\theta)=\frac{m_y/v-p_z\sin\theta}{\varepsilon_0\varepsilon_{\rm sur}E_0}\:.
\end{equation}
The reflectance is determined by two SLRs related to a strong, in-plane ED excitation and a relatively weak, in-plane MD one (Fig.~S5(a2--d2)). 
Similarly to what was observed for the TE-polarised illumination, the reflectance shows a dip (Fig.~\ref{fig:r_glycerin_num}(a3)) originating from the destructive interference in the backward direction between the ED and MD components $p_x$ and $m_y$ (the so-called Kerker effect \cite{babicheva2024mie}). 
In contrast to the TE case, a weak coupling between quasi-sBIC with $p_z$ and the SLR with $m_y$ is realized at small angles of incidence, as the spectral overlap between these modes is smaller than their linewidth at normal incidence. 
As the angle of incidence increases, a hybrid mode with the participation of $p_z$ and $m_y$ is formed along the Rayleigh anomaly.

Interestingly, at higher angles of incidence ($\theta\gg1^{\circ}$), the TM polarised excitation leads to the in-plane MD component $m_y$ and the out-of-plane ED component $p_z$ with comparable magnitude in the hybrid mode. 
Their coupling, therefore, allows for the realization of a condition of perfect annihilation between the far-field radiation of the quasi-sBIC part (induced by $p_z$) and that of the SLR part (induced by $m_y$) of the hybrid mode (Fig.~S6(a2,b2,c)), perfectly cancelling each other out: an accidental BIC (aBIC)~\cite{abujetas2020coupled,abujetas2022tailoring,Yingying2024}. 
This feature has been measured (Fig.~\ref{fig:t_g_exp}) and found analytically (Fig.~\ref{fig:r_glycerin_num}(a3)) for an angle of incidence $\theta\approx 20^{\circ}$. The appearance of the aBIC can be seen as a result of the self-adjustment of the contributing dipole moment components governed by the coupling strength (cf. TE and TM panels in Fig.~\ref{fig:r_glycerin_num}). A similar feature is not observed in the TE polarisation case for the metasurface considered here, because the in-plane and out-of-plane dipolar components exhibit considerably unbalanced magnitudes.
It is in principle possible to realize an aBIC with both TE and TM polarisations at the same wavelength and angle via a fine-tuning of the rectangular lattice and particle parameters~\cite{abujetas2022tailoring}. 

As already noted, the formation of the collective modes of the metasurface is accompanied by a marked enhancement of the near field, with its spatial distributions depending strongly on the polarisation of the illumination. In the case of TM polarisation, the strong electric field is concentrated between the meta-atoms of the metasurface, whereas for TE polarisation, the electric field is enhanced in the vicinity of the meta-atoms (indicated by arrows 3 and 7, respectively, in Fig.~S4). For practical purposes, this can be used for selective excitation and switching of active functions, for example, quantum dots deposited on the metasurface. It is also interesting to note that for both cases of the aBIC (TM polarisation, fields 1-3 in Fig.~S4) and the anticrossing due to strong coupling (TE polarisation, fields 4-6 in Fig.~S4), the electric fields are considerably reduced, leading to the suppression of reflection.

\subsection{Tuning strong coupling with the metasurface environment}

\begin{figure*}[!b]
\centering
\includegraphics[width=1\linewidth]{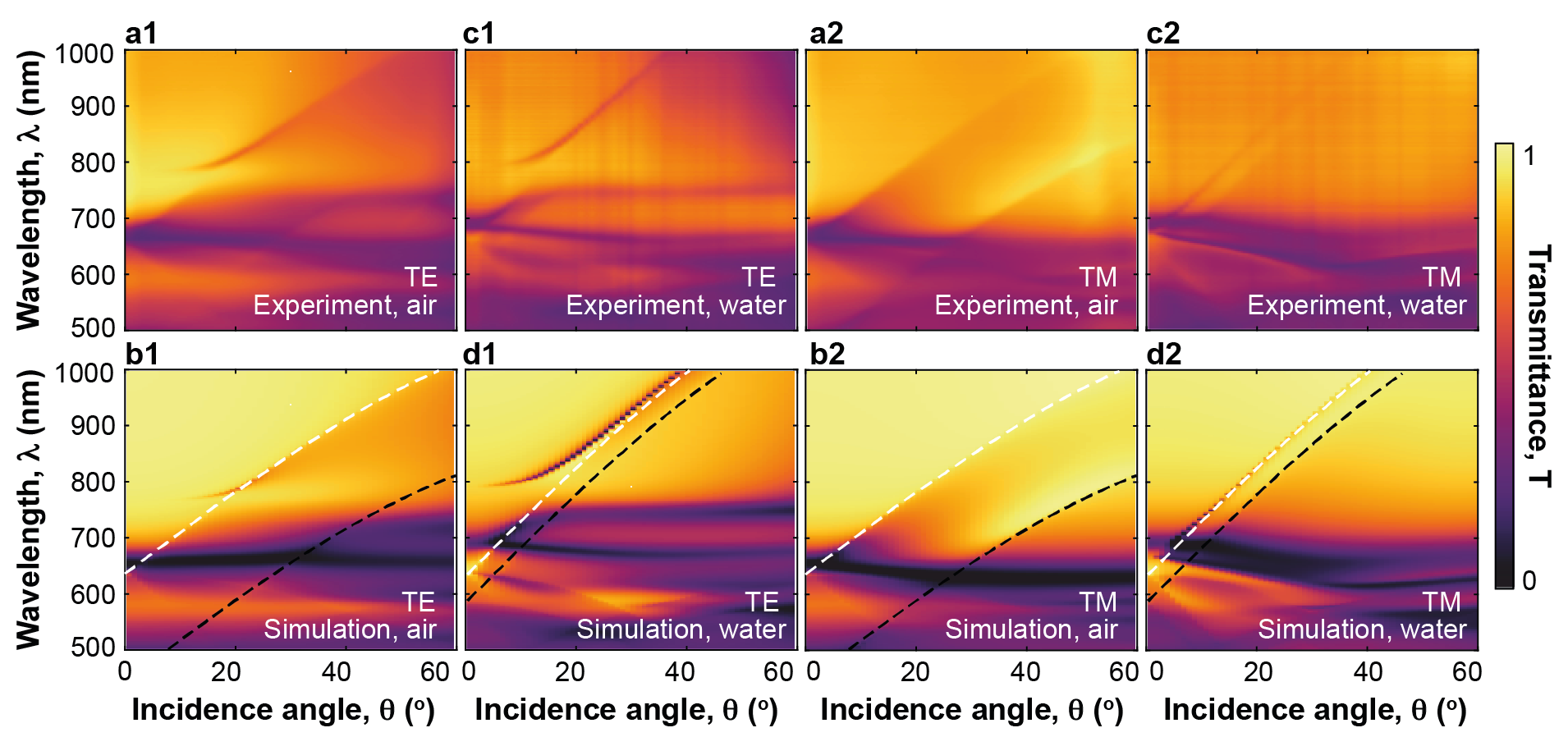}
\caption{Experimental (top) and numerical (bottom) transmittance spectra of the measurface with (a,b) air and (c,d) water superstrate for different incident polar angles $\theta$ for (1) TE- and (2) TM-polarised illumination. Dashed lines indicate the main-order Rayleigh anomaly for (white) the substrate and (black) the superstrate. The values of the wavelength $\lambda$
 are indicated for vacuum.}
\label{fig:t_air_water}
\end{figure*}
The response of the metasurface to light is expected to show a strong dependence on the refractive index of the surrounding medium. In particular, with the Si nanodisks on a glass substrate, this dependence translates into a dependence on the refractive index mismatch between the substrate (glass) and the medium used as a superstrate. Submerging the metasurface in glycerol provides practically uniform surrounding eliminating the mismatch. This is a well-known requirement for the existence of collective resonances~\cite{Allayarov2023_APR}. To explore the possibility of tuning the strong coupling mechanism by varying the surrounding environment, air and water superstrates were investigated (Fig.~\ref{fig:t_air_water}(a1,a2,c1,c2)) in addition to glycerol (Fig.~\ref{fig:t_g_exp}). 
The analytical CDM introduced above cannot be directly applied in the case of an inhomogeneous environment, as it does not take into account the dipole radiation reflections at the interface between substrate and superstrate. 

The experimental reflectance spectra and their numerical simulation for different superstrates are shown in Fig.~S7 of the Supporting Information.
For TE polarisation, the anticrossing character of the resonances is observed even in the presence of a dielectric contrast between the substrate and the superstrate (Fig.~\ref{fig:t_air_water}(a1,c1)). This indicates the preservation of the strong coupling regime between the corresponding collective resonances, with a reduction of the splitting for increasing mismatch. For example, the estimated splitting for air, water, and glycerol superstrates is 30~nm, 60~nm and 65~nm, respectively (more details on the physical mechanism behind it can be found in Supplementary Information and are illustrated in Fig.~S8).
For TM polarisation, the spectral features associated with the superstrate undergo a blue-shift, and several resonant phenomena are considerably damped. In particular, the progressive separation between the sBIC and SLR with the increasing index mismatch leads to the complete loss of the aBIC feature, which was observed with glycerol (Fig.~\ref{fig:t_g_exp}(c1)). Therefore, the observed phenomena suggest strategies to tune the strength of the coupling between collective resonances via dynamic modification of the external environment.

\section{Conclusion}
We investigated coupling between nonlocal collective resonances of a dielectric metasurface supporting excitation of SLR and BIC modes. Theoretical analysis and experimental observations reveal the emergence of either strong or weak coupling between SLRs and quasi-sBICs. This leads to the formation of hybrid resonances consisting of coupled modes, manifesting as anticrossing (or Rabi splitting) and aBIC under TE and TM excitation, respectively. 
Since the metasurface building block mainly exhibits dipolar response in the considered spectral range, we applied the coupled-dipole model to describe optical properties under oblique illumination. This model can describe the formation of SLR and sBIC modes, which are associated with in-plane and out-of-plane electric (or magnetic) dipoles, respectively. While such modes are independent at normal incidence, the excitation of both in-plane and out-of-plane components at oblique illumination results in their coupling.
Important for potential applications, we have numerically and experimentally shown that the strong coupling can be tuned by varying the superstrate material, and its activation is possible simply via polarisation control. We believe that the obtained results provide a framework for understanding the physics behind dipole-dipole interactions leading to the hybridization of collective modes, and are important for practical applications in sensing, nonlinear optics, enhancement and control of light-matter interaction, tunable nanophotonics, and single and entangled photon generation.

\section{Methods}
\subsection{Spectroscopic measurements}
\begin{figure}[!b]
\centering
\includegraphics[scale = 1]{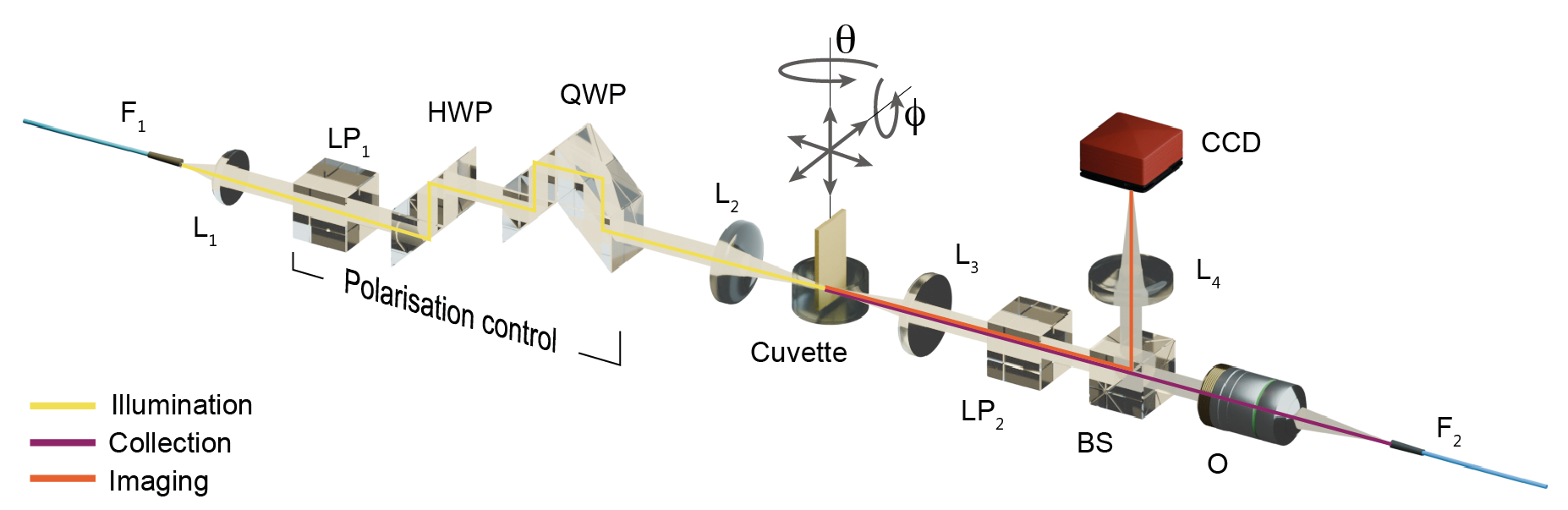}
\caption[Experimental Setup]{Schematics of the experimental setup for spectroscopic measurements. The optical components are labelled as follows: F$_{1,2}$ are the illumination and collection fibres, respectively, L$_i$ are the achromatic doublets of focal lengths $f_1$ = 35~mm, $f_{2,3}$ = 75~mm, $f_4$ = 200~mm, LP$_i$ are the linear polarisers, HWP and QWP are the half- and quarter-waveplates, respectively, BS is the 50/50 beam splitter cube, O is the collection objective (NA = 0.42, WD = 4.5~mm). The metasurface can be submerged in the cuvette to perform the measurements in different environments (air, water, glycerol). Grey arrows around the sample show its spatial degrees of freedom. (Representation of optical components in this schematics were designed by Ryo Mizuta Graphics). }
\label{fig:exp_setup}
\end{figure}
The transmission and reflection spectra were measured for different angles of incidence and polarisations of the illuminating light, as well as with various superstrates (air, de-ionised water and pure glycerol). The system consists of a tabletop custom-built spectroscopy setup (Fig.~\ref{fig:exp_setup}), that can be divided into three sections built around the sample: the illumination, the imaging and the collection branches. The latter two are always kept at the same position, while the illumination branch and the sample can be independently rotated around a common axis by controlling two motors via custom-made software written in Python. After an initial careful alignment of the sample to the system rotation axis, both transmission and reflection spectra of the sample can be measured with great control over the angle of incidence. In the transmission configuration the source is kept at the position represented in Fig.~\ref{fig:exp_setup} and the sample is rotated about the central axis therein shown. For measuring reflection, the source and the sample are both rotated about the central axis ensuring that light reflected by the sample is collected through the detection branch.

A broadband halogen lamp (HL-2000-HP, Ocean Optics) is fibre-coupled ($F_1$) and collimated by a short lens (L$_1$). The source is mounted on the illumination branch to allow rigid rotation with it. After collimation, a series of three prisms is used for full and broadband control of the state of polarisation of the illuminating light. A Glan-Taylor prism initially fixes the polarisation to the horizontal direction (parallel to the optical table). A double and a single Fresnel rhombs are then used to modify the polarisation as required. These prisms work respectively as a broadband half- and quarter-waveplate allowing to obtain linear, elliptical or circular light, with any orientation or helicity. For the measurements reported here, the illumination polarisation has been fixed to right-handed circular, and a second Glan-Taylor prism is used to select the polarisation of interest in the readout as either TE or TM. Finally, light is focused on the sample via $L_2$. The location of the collimating lens $L_1$ is fixed so that a broad and uniform illumination of the sample is obtained.

A lens matching the focal length of $L_3$ is used to collect light transmitted (or reflected) by the sample, which is subsequently sent to both the imaging and collection branches, via a beam splitter cube. The surface of the sample is then imaged onto a camera via $L_5$ (f = 200~mm), and on the collection fibre ($F_2$) via a microscope objective $O_1$ (NA = 0.42, WD = 45~mm). The fibre is in turn coupled to a spectrometer.

To change the metasurface surroundings, the sample mount consists of two sections. Vinculated to the main rotation motor, a post mounting a round plate hosts a glass cylindrical cuvette, whose height can be adjusted to meet the light beam. A pair of post holders are mounted on the side of the same rotation motor to hold a breadboard above the glass cuvette via two long posts. The sample holder is mounted upside down on said breadboard so that it can be inserted from the top in the cuvette volume, once this has been filled with the medium of interest. For a finer adjustment of the sample position and orientation, the sample mount allows three-dimensional translation and rotation of both azimuth and tilt angles.

\subsection{Numerical simulations}
Numerical simulations were performed using the ANSYS Lumerical software. The properties of a single Si disk such as scattering efficiency, dipole moments and polarizabilities were calculated with the FDTD solver using a simulation domain of $800\times800\times800$~nm$^3$, boundary conditions of 32 layers of standard PML in all directions, mesh accuracy of 8, mesh refinement of conformal variant 1. For more details, please see Methods section of Ref.~\cite{Allayarov2023_APR} and Appendices of Ref.~\cite{Allayarov2024_OE}.
The reflectance and transmittance of a periodic array of disks were calculated by using the RCWA solver with a circular k-vector domain, max number of k-vectors equal to 225, tangent vector field activated and mesh refinement of conformal variant 1.

\backmatter

\bmhead{Supplementary information}

The paper is supported by the Supplementary Information (see below).

\bmhead{Acknowledgements}
We acknowledge the central computing cluster operated by Leibniz University IT Services (LUIS), which is funded by the DFG (project number INST 187/742-1 FUGG) and Cornerstone for their help in the fabrication of the samples. We sincerely thank Prof. Andrei Laurynenka and Dr. Radu Malureanu from Technical University of Denmark (DTU) for providing SOI wafers.


\section*{Funding}
This work was supported by Deutsche Forschungsgemeinschaft (DFG, German Research Foundation) under Germany’s Excellence Strategy within the Cluster of Excellence PhoenixD (EXC 2122, Project ID 390833453), the Alexander von Humboldt Foundation, the ERC iCOMM project (789340), and the UK EPSRC project EP/Y015673/1.







\section*{Author contribution}
I.A. performed numerical simulations. V.A., D.J.R., B.v.C. and A.Y.B performed experiments. I.A. and A.B.E. performed analytical studies. I.A., V.A., A.B.E., A.V.Z., and A.C.L. contributed to formulating the research direction and discussing the results. I.A., V.A., A.B.E., and A.C.L. wrote the first draft. All authors revised the paper. A.B.E., A.V.Z., and A.C.L. provided supervision.


\bibliography{references}

\end{document}


\title[Article Title]{Supporting Information: Strong coupling of collective optical resonances in dielectric metasurfaces}


\author[1,2,3]{\fnm{Izzatjon} \sur{Allayarov}}

\author[4]{\fnm{Vittorio} \sur{Aita}}

\author[4,5]{\fnm{Diane J.} \sur{Roth}}

\author[4]{\fnm{Boaz} \sur{van Casteren}}

\author[4,6]{\fnm{Anton Yu.} \sur{Bykov}}

\author*[3,7]{\fnm{Andrey B.} \sur{Evlyukhin}}\email{evlyukhin@iqo.uni-hannover.de}

\author*[4]{\fnm{Anatoly V.} \sur{Zayats}}\email{anatoly.zayats@kcl.ac.uk}

\author*[1,2,3]{\fnm{Antonio} \sur{Cal{\`a} Lesina}}\email{antonio.calalesina@hot.uni-hannover.de}

\affil*[1]{\orgdiv{Hannover Centre for Optical Technologies}, \orgname{Leibniz University Hannover}, \orgaddress{\street{Nienburger Str. 17}, \city{Hannover}, \postcode{30167}, \state{State}, \country{Germany}}}

\affil[2]{\orgdiv{Institute of Transport and Automation Technology}, \orgname{Leibniz University Hannover}, \orgaddress{\street{An der Universit{\"a}t 2}, \city{Garbsen}, \postcode{30823}, \country{Germany}}}

\affil[3]{\orgdiv{Cluster of Excellence PhoenixD}, \orgname{Leibniz University Hannover}, \orgaddress{\street{Welfengarten 1A}, \city{Hannover}, \postcode{30167}, \country{Germany}}}

\affil[4]{\orgdiv{Department of Physics and London Centre for Nanotechnology}, \orgname{King's College London}, \orgaddress{\street{Strand}, \city{London}, \postcode{WC2R 2LS}, \country{UK}}}

\affil[5]{\orgdiv{QinetiQ}, \orgname{Cody Technology Park}, \orgaddress{\street{Ively Road}, \city{Farnborough}, \postcode{GU14 0LS}, \country{UK}}}

\affil[6]{\orgdiv{NanoPhotonics Centre, Cavendish Laboratory, Department of Physics}, \orgname{University of Cambridge}, \orgaddress{\street{}, \city{Cambridge}, \postcode{CB3 0HE}, \country{UK}}}

\affil[7]{\orgdiv{Institute of Quantum Optics}, \orgname{Leibniz University Hannover}, \orgaddress{\street{Welfengarten 1}, \city{Hannover}, \postcode{30167}, \country{Germany}}}




\maketitle

\section{Lattice sums}

The explicit forms of the lattice sums ($S_x$, $S_y$, $S_z$) and the coupling parameter $g_x$ can be written as~\cite{allayarov2025analytical}
\begin{align}
     &g_x=-k_{\rm sur}\sum_{n=0}^{\infty}\sum_{m=0}^{\infty} x_n F_{nm} \left(\frac{{\rm i}k_{\rm sur}}{r_{nm}}-\frac{1}{r_{nm}^2}\right) \ne 0\quad {\rm if}\quad {\sin\theta}\ne 0,\\
     &S_{x}=k_{\rm sur}^2\sum_{n=0}^{\infty}\sum_{m=0}^{\infty}F_{nm}\left(1+\frac{{\rm i}}{k_{\rm sur}r_{nm}}-\frac{1}{k_{\rm sur}^2r_{nm}^2}-\frac{x_n^2}{r_{nm}^2}-\frac{3{\rm i}x_n^2}{k_{\rm sur}r_{nm}^3}+\frac{3x_n^2}{k_{\rm sur}^2r_{nm}^4}\right),\\
     &S_{y}=k_{\rm sur}^2\sum_{n=0}^{\infty}\sum_{m=0}^{\infty}F_{nm}\left(1+\frac{{\rm i}}{k_{\rm sur}r_{nm}}-\frac{1}{k_{\rm sur}^2r_{nm}^2}-\frac{y_m^2}{r_{nm}^2}-\frac{3{\rm i}y_m^2}{k_{\rm sur}r_{nm}^3}+\frac{3y_m^2}{k_{\rm sur}^2r_{nm}^4}\right),\\
     &S_{z}=k_{\rm sur}^2\sum_{n=0}^{\infty}\sum_{m=0}^{\infty}F_{nm}\left(1+\frac{{\rm i}}{k_{\rm sur}r_{nm}}-\frac{1}{k_{\rm sur}^2r_{nm}^2}\right),
\end{align}
where $k_{\rm sur}$ is the wave number in a medium with a refractive index of $n_{\rm sur}$, $r_{nm}=|{\bf r}_{nm}|=\sqrt{x_n^2+y_m^2}$, $x_n = Pn$, $y_m = Pm$ and $F_{nm} = {\rm exp}[{\rm i}k_{\rm sur}(r_{nm}+x_n\sin\theta)]/(4\pi r_{nm})$, $n$ and $m$ are integer numbers, $P$ is the period of a square lattice, $\theta$ is the polar angle of the of incidence. Note that $n=m=0$ term must be excluded from the above sums. For the sake of the convenience, we assume that all particles are placed on the $xy$-plane at $z=0$. 


\section{Supplementary figures}
\begin{figure*}[!htb]
\centering
\includegraphics[width=1\linewidth]{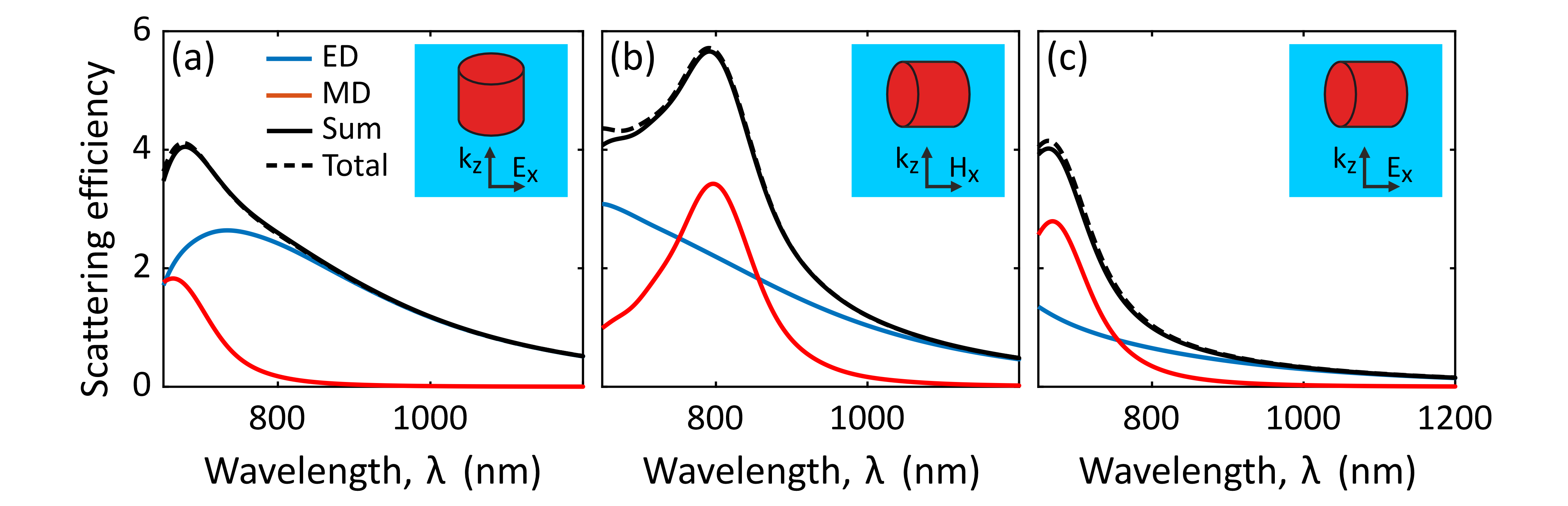}
\caption{Simulated scattering efficiency of a single polycrystalline Si nanodisk for different excitation conditions indicated in the insets. The individual disk has a diameter $D=220$~nm, height $H=100$~nm and a refractive index of surroundings $n_{\rm sur}$=1.45. ED (blue): electric dipole contribution, MD (red): magnetic dipole contribution, Sum: ED + MD, Total: all multipoles. Comparison of Sum and Total shows that only dipolar response of the nanodisk is important in the considered spectral range.}
\label{fig:single}
\end{figure*} 

\begin{figure*}[!htb]
\centering
\includegraphics[width=1\linewidth]{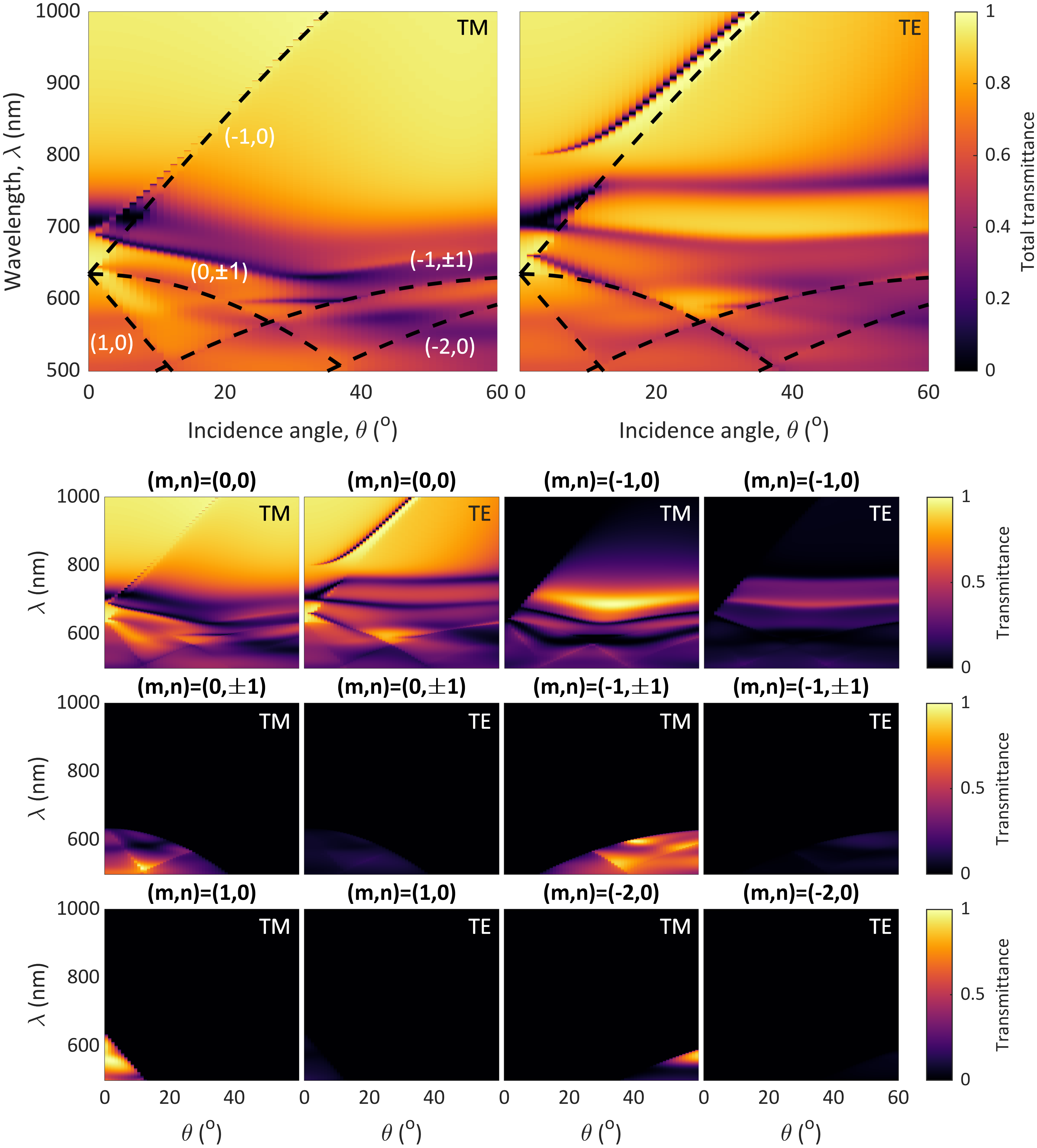}
\caption{Total transmittance (top row panels) of the metasurface considered in Fig.~2 of the main text for TM and TE polarised illumination. The black dashed lines indicate the  positions of all Rayleigh anomaly orders (orders are given in brackets and the same for both TM and TE cases) for the considered incidence angle and wavelength range. The panels below present the decomposition of the total transmittance into the grating order contributions.}
\label{fig:RAdec}
\end{figure*}

\begin{figure*}[!htb]
\centering
\includegraphics[width=1\linewidth]{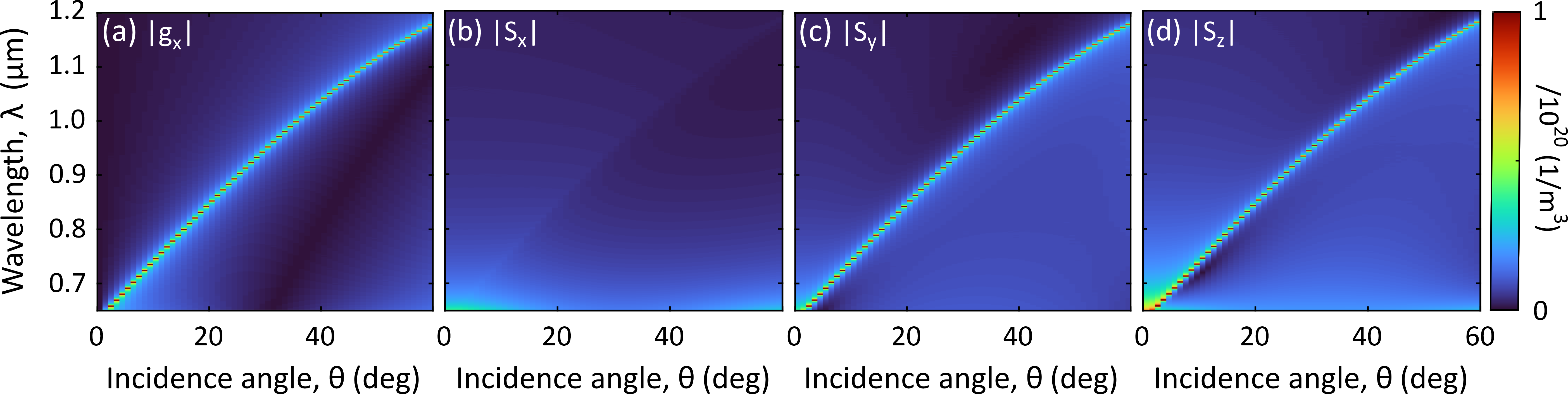}
\caption{Simulated dispersions with respect to polar angle of incidence $\theta$ of the absolute value of (a) the coupling parameter $g_x$ and the dipole lattice sums (b) $S_x$, (c) $S_y$, and (d) $S_z$. The period of a square lattice is $P=440$~nm and the refractive index of the environment is $n_{\rm sur}$=1.45.}
\label{fig:sg}
\end{figure*}

\begin{figure*}[!htb]
\centering
\includegraphics[width=1\linewidth]{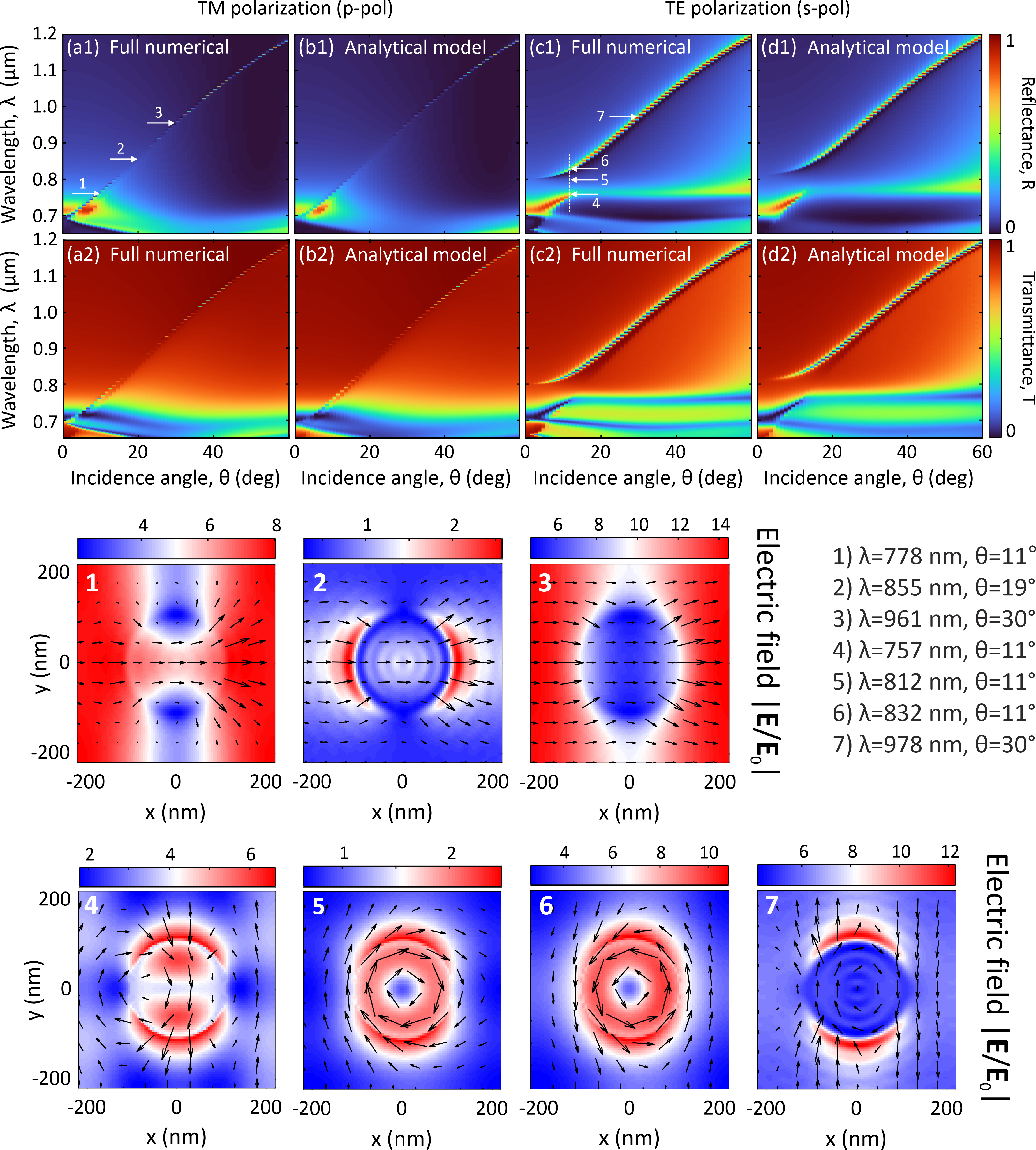}
\caption{Comparison of the theoretical results from the full numerical and the coupled-dipole model simulations for the reflectance and transmittance dispersion of the metasurface immersed in a homogeneous environment ($n_{\rm sur}=n_{\rm sub}=n_{\rm sup}$=1.45) for (a,b) TM- and (c,d) TE-polarised illumination. Only the main diffraction orders are shown. The bottom panels show the normalized electric field distributions at the positions indicated by numbers in panels (a1) and (c1). $E_0$ is the electric field of the incident wave. The white dashed circle indicates a Si nanodisk. The metasurface with a period $P=440$~nm consists of nanodisks with a diameter $D=220$~nm, height $H=100$~nm. Surrounding refractive index is $n_{\rm sur}$=1.45.}
\label{fig:an_vs_num}
\end{figure*}

\begin{figure*}[!htb]
\centering
\includegraphics[width=0.85\linewidth]{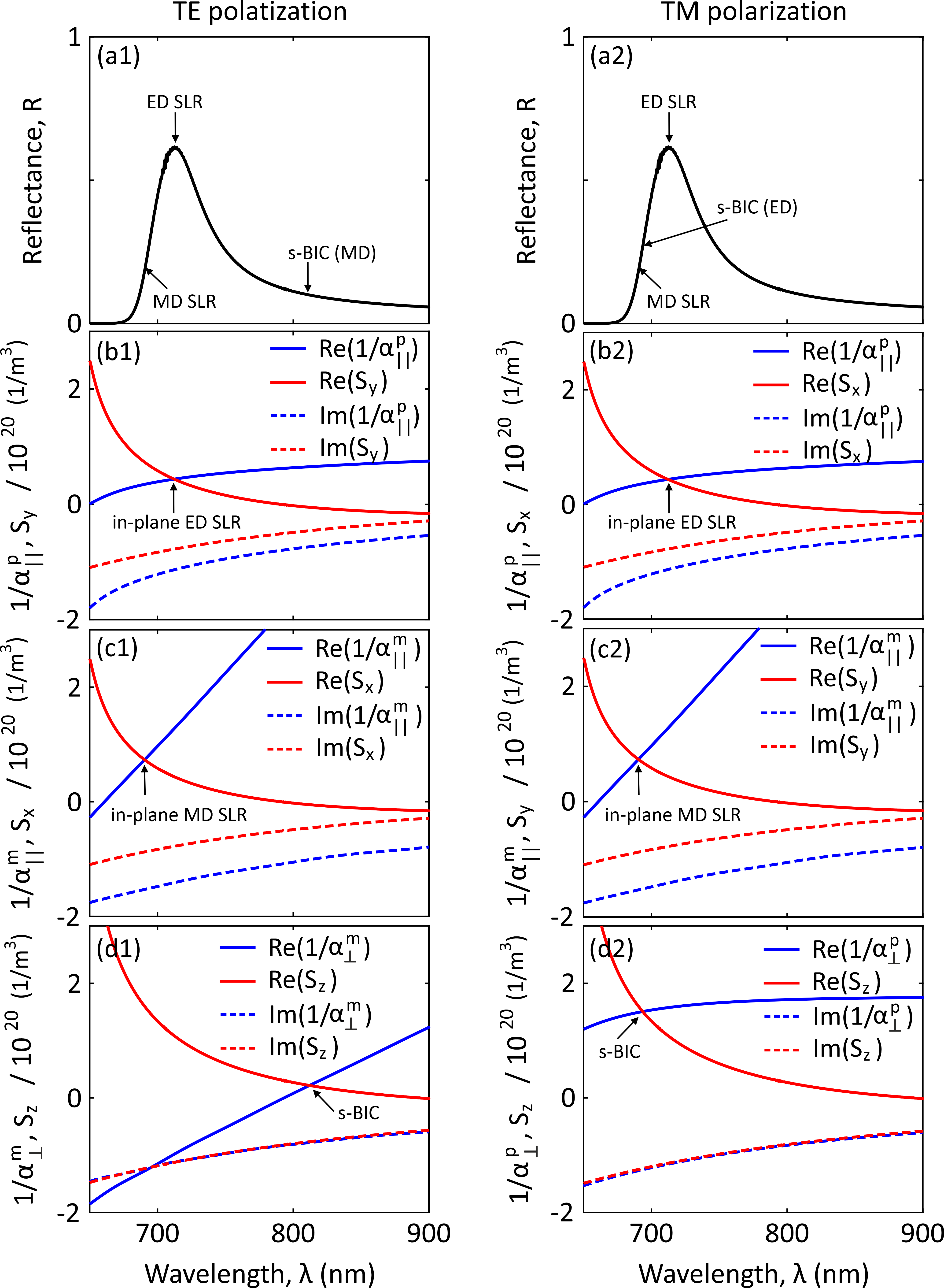}
\caption{(a) Simulated total reflectance with the indicated positions of the SLR and the sBIC for (1) TE- and (2) TM-polarised illumination at normal incidence ($\theta$=0). (b--d) The conditions leading to the spectral features associated with (b) in-plane ED SLR, (c) in-plane MD SLR, and (d) symmetry-protected BIC.}
\label{fig:theta0}
\end{figure*}

\begin{figure*}[!htb]
\centering
\includegraphics[width=1\linewidth]{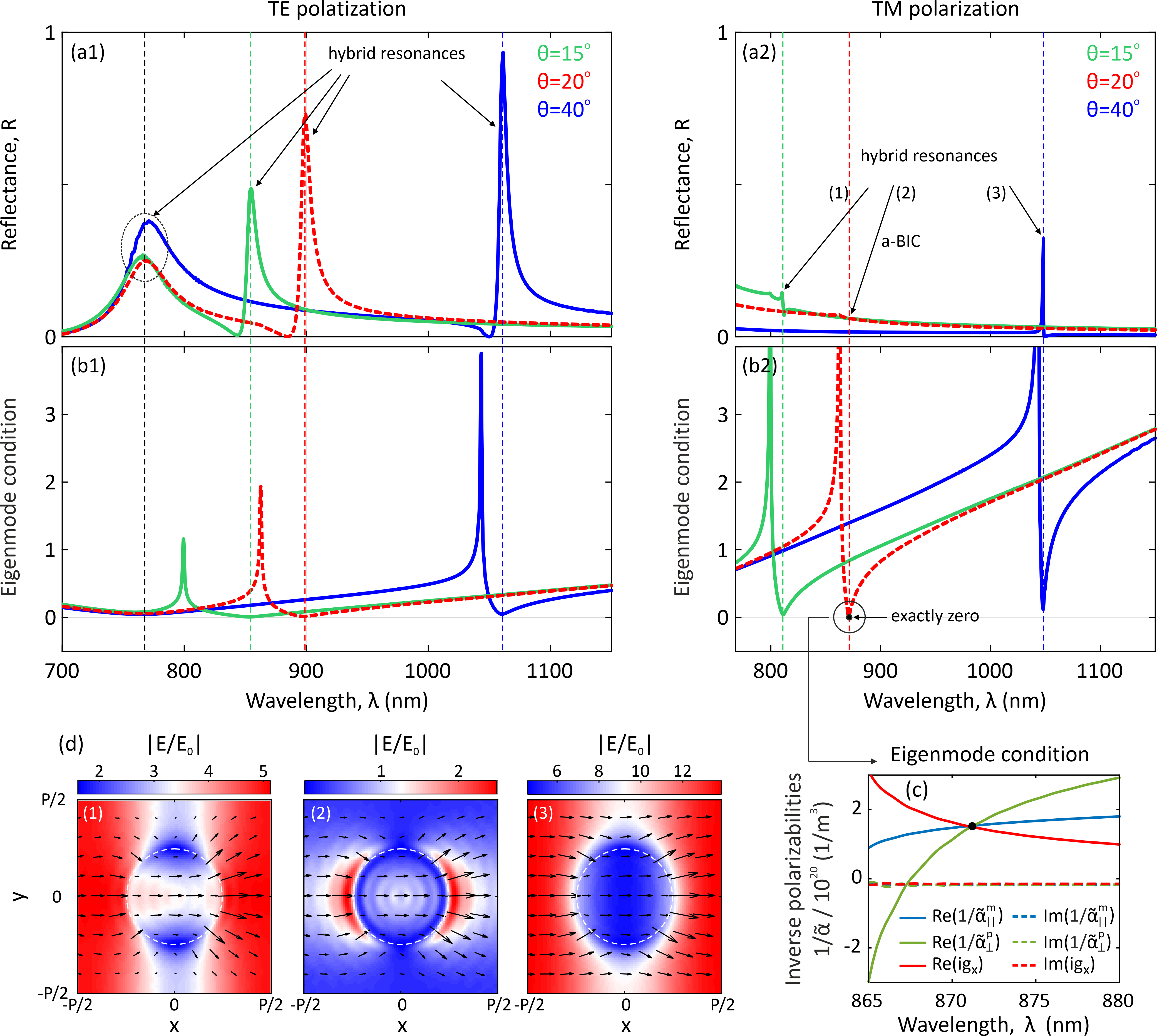}
\caption{(a) Reflectance and (b) eigenmode conditions at three different angles of incidence for (1) TE- and (2) TM-polarised excitation. The eigenmode conditions are $({1}/{\alpha^{\rm p}_{\parallel}}-S_y)({1}/{\alpha^{\rm m}_{\perp}}-S_z)+g_x^2 = 0$ for TE, and $({1}/{\alpha^{\rm m}_{\parallel}}-S_y)({1}/{\alpha^{\rm p}_{\perp}}-S_z)+g_x^2 = 0$ for TM case. The reflectance resonances corresponds to the minima of the eigenmode conditions but have non-zero bandwidth since the conditions are not satisfied (i.e., $\neq$0). When they are satisfied, e.g., for $\theta=20^\circ$ in the panel (b2), the resonant state becomes a nonradiative eigenstate [also known as an accidental BIC (aBIC)] of the system with zero bandwidth (and infinite Q-factor). (c) Visualization of the condition for aBIC, i.e., $1/\tilde{\alpha}^{\rm m}_{||}=1/\tilde{\alpha}^{\rm p}_{\perp}=ig_x$, where $1/\tilde{\alpha}^{\rm m}_{||}=(1/{\alpha}^{\rm m}_{||}-S_y)\sin\theta$ and $1/\tilde{\alpha}^{\rm p}_{\perp}=(1/{\alpha}^{\rm p}_{\perp}-S_z)/\sin\theta$~\cite{abujetas2022tailoring,allayarov2025analytical}. Under these conditions, the in-plane MD and out-of-plane ED radiation interferes destructively in the far-field region completely cancelling each other out. (d) The field distributions at the wavelengths indicated by the arrows in the panel (a2). Accidental BIC cannot be excited by an external field of a plane wave as one can see from the field distributions in (d), i.e., the electric field is much smaller at the condition of the aBIC [see (2) in (d)] than quasi-aBIC (excited) cases [see (1) and (3) in (d)].}
\label{fig:rescond}
\end{figure*}

\begin{figure*}[!htb]
\centering
\includegraphics[width=1\linewidth]{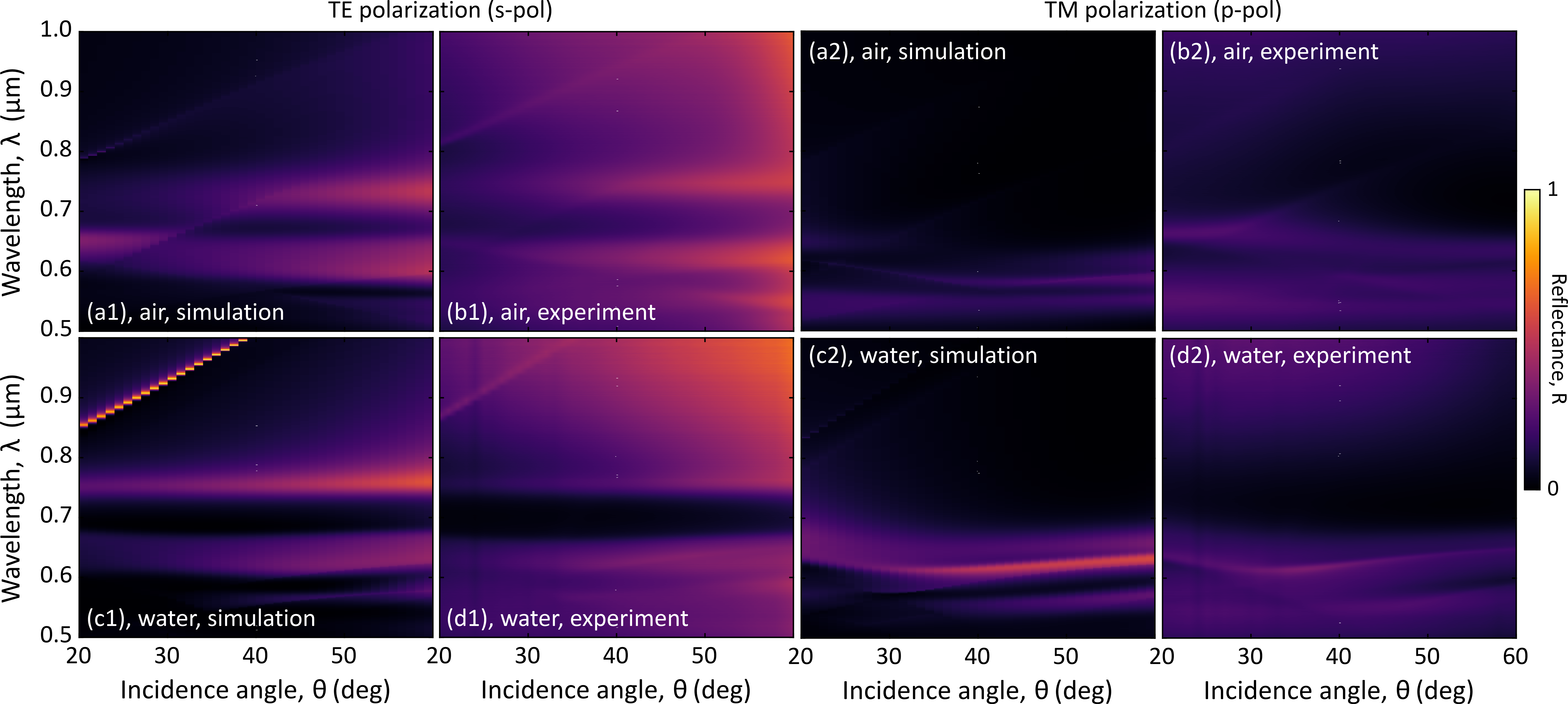}
\caption{Comparison of (a,c) simulated and (b,d) experimental reflectance dispersion of the metasurface on a glass substrate in the case of (top) air and (bottom) water superstrate for (1) TE and (2) TM polarisations.}
\label{fig:r_aw}
\end{figure*}

\begin{figure*}[!htb]
\centering
\includegraphics[width=0.85\linewidth]{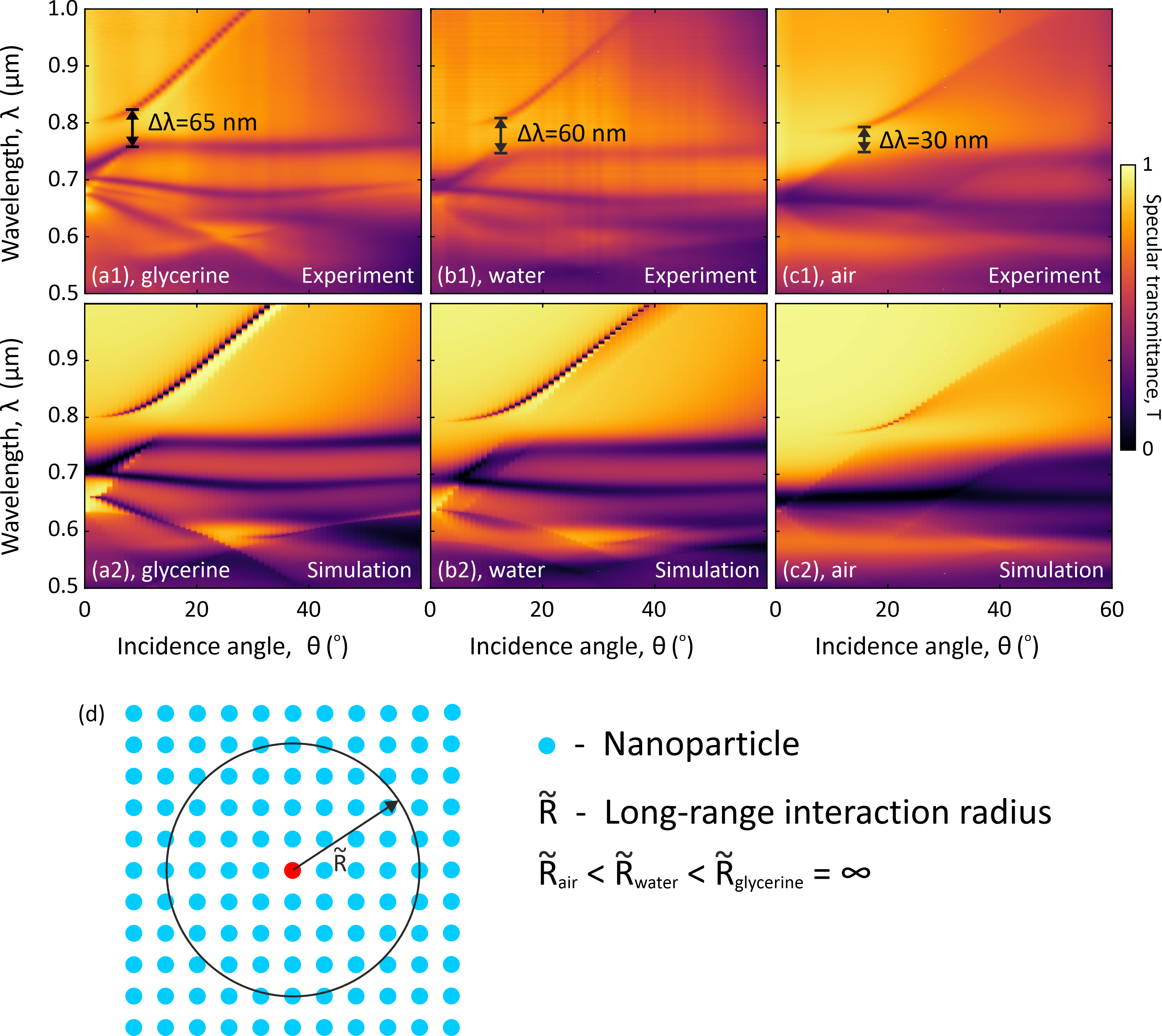}
\caption{Measured (top row) and simulated (bottom row) specular transmittance of the metasurface on a glass substrate and with (a) glycerine, (b) water and (c) air superstrate for the TE polarised illumination. $\Delta\lambda$ is the anticrossing gap, which shrinks with the increase of the substrate-to-superstrate index contrast $\Delta n = n_{\rm sub} - n_{\rm sup}$. This can be explained considering the lattice sums $S$ associated with the electromagnetic interaction between metasurface dipoles of the same order, which can be represented as a sum of near-, middle-, and far-field contributions. The far-field part, in contrast to the other two, diverges at the Rayleigh anomaly wavelength ($\lambda^{\rm RA}$) in the case of homogeneous surroundings ($\Delta n = 0$). For inhomogeneous surroundings ($\Delta n \neq 0$), the far-field part of S becomes finite at $\lambda^{\rm RA}$, as has been shown in Ref.~\cite{Allayarov2023_APR}. This is due to the far-field interaction between the nanoparticles being suppressed as a consequence of reflection at the substrate-superstrate interface. (d) Schematic of an array of nanoparticles placed on a glass substrate, indicating the long-range interaction radius $\Tilde{R}$ between the nanoparticles, which is inversely proportional to $\Delta n$~\cite{Allayarov2023_APR}. For the glycerine superstrate (homogeneous surrounding), $\Tilde{R}=\infty$ and, therefore, each particle interacts with all particles of the metasurface. For the cases of water and air superstrate, $\Tilde{R}$ is finite and each particle interacts with a finite number of particles around it. The splitting gap $\Delta\lambda$ becomes smaller for a higher index contrast $\Delta n$ since a limited number of particles participate in the interaction, resulting in a weaker coupling than in homogeneous surroundings.} 
\label{fig:r_gwa}
\end{figure*}

\clearpage


\bibliography{references}